# High-Q Tantalum Oxide Nanomechanical Resonators by Laser-Oxidation of $TaSe_2$

Santiago J. Cartamil-Bueno[1] (✉), Peter G. Steeneken[1] (✉), Frans D. Tichelaar[2], Efren Navarro-Moratalla[3], Warner J. Venstra[1], Ronald van Leeuwen[1], Eugenio Coronado[3], Herre S.J. van der Zant[1], Gary A. Steele[1], and Andres Castellanos-Gomez[1,†] (✉)

[1] Kavli Institute of Nanoscience, Delft University of Technology, Lorentzweg 1, 2628 CJ Delft, The Netherlands

[2] Kavli Institute of Nanoscience, Delft University of Technology, National Centre for HREM, Lorentzweg 1, 2628 CJ Delft, The Netherlands

[3] Instituto Ciencia Molecular (ICMol), Univ. Valencia, C/Catedrático José Beltrán 2, E-46980, Paterna, Spain

[†] Instituto Madrileño de Estudios Avanzados en Nanociencia (IMDEA-Nanociencia), 28049 Madrid, Spain

## ABSTRACT

Controlling the strain in two-dimensional materials is an interesting avenue to tailor the mechanical properties of nanoelectromechanical systems. Here we demonstrate a technique to fabricate ultrathin tantalum oxide nanomechanical resonators with large stress by laser-oxidation of nano-drumhead resonators made out of tantalum diselenide ($TaSe_2$), a layered 2D material belonging to the metal dichalcogenides. Prior to the study of their mechanical properties with a laser interferometer, we checked the oxidation and crystallinity of the freely-suspended tantalum oxide in a high-resolution electron microscope. We show that the stress of tantalum oxide resonators increase by 140 MPa (with respect to pristine $TaSe_2$ resonators) which causes an enhancement of quality factor (14 times larger) and resonance frequency (9 times larger) of these resonators.

### Introduction

Two-dimensional layered materials are attractive for high-frequency nanomechanical systems, which can be used in sensing applications. The reduced thickness and small mass of these materials enable high resonance frequencies $f_0$ and fast response times, whereas their low flexural rigidity increases the responsivity and allows size reduction. For low-noise operation of nanomechanical systems it is desirable to achieve high quality factors $Q$ at high frequencies. In conventional nanomechanical systems, based on silicon nitride ($Si_3N_4$) beams, it has been shown that both $f_0$ and $Q$ can be enhanced by increasing the stress in the beam [1, 2]. For this purpose, several methods have been proposed to tune the stress in nanomechanical systems based on 2D materials using temperature, mechanical actuators and gas pressure [3–6]. For permanent stress modification in polycrystalline graphene, a method for direct bonding between graphene platelets has been proposed [6]. Since the mechanical properties of suspended crystalline 2D materials attract much attention for sensing applications, it is of scientific and technological interest to develop methods for local stress engineering in single-crystalline 2D materials.

In this work we report a method that permanently modifies the stress in suspended single-crystalline 2D materials. We use a focused laser to locally increase the temperature of tantalum diselenide ($TaSe_2$) drumhead resonators until they undergo an oxidation process which results in a drastic increase in the pre-stress of the selected 2D resonators. We investigate the effect of this increased pre-stress on the Q factor and resonance frequency, and find a large enhancement for both of them in thin resonators (< 20 nm thick). It thereby provides a route towards higher $f$-$Q$ products in layered material resonators. The stress and thickness dependence of the Q factor are observed to be governed by the same model that was proposed to describe the Q factor of stressed $Si_3N_4$ beams [1, 2, 7].

## Experimental

### Fabrication of $TaSe_2$ drumhead resonators

$TaSe_2$ flakes are exfoliated from synthetic $TaSe_2$ crystals by mechanical exfoliation with blue Nitto tape (Nitto Denko Co., SPV 224P) [8]. More details can be found in Section I and II in the Electronic Supplementary Material (ESM).

### Laser-oxidation of $TaSe_2$

A Renishaw *in via* system is used to scan a focused laser spot ($\lambda$ = 514 nm) over the $TaSe_2$ flakes. The oxidation of material is found to occur at a power of 25 mW for 0.1-0.2 s of exposure time. The scanning step in the irradiation process is 300 nm.

### Material characterization

Micro-Raman spectrometry and photoluminescence are used to characterize microregions of the materials, and a high-resolution electron microscope (Tecnai F20ST/STEM) with EDX detector allowed atomic resolution imaging and analysis of the elemental composition. See Section IV and V in the ESM.

### Mechanical resonance measurement

The frequency response of the drumhead resonators is obtained by means of optical interferometry [9, 10]. We analyze the fundamental mode, which is easily identified as it shows the lowest frequency and highest intensity among all the mechanical resonance peaks in the spectrum. The measurement is carried out in vacuum (~$10^{-5}$ mbar) at room temperature. More information is available in Section VI and IX in the ESM.

**Results and discussion**

For Transmission Electron Microscopy (TEM) characterization, mechanically exfoliated $TaSe_2$ flakes are deposited onto a 200 nm $Si_3N_4$ membrane with holes (2.5 μm in diameter), using a recently developed dry transfer technique [11]. A focused green laser is scanned over part of the flakes to induce the local laser-oxidation in a confocal microscope system operated in air. Figure 1(a) shows a transmission mode optical image of the partially oxidized 50 nm thick $TaSe_2$ flake. The regions labelled with "1", "2" and "3" in Fig. 1(a) correspond to the pristine flake, laser-exposed flake and bare $Si_3N_4$ membrane, respectively. A dramatic difference in the optical properties is observed between pristine and laser-exposed areas of the flake even though the change in thickness of the laser-irradiated flakes was small (~3 nm), as found from atomic force microscopy (AFM, see Section III in the ESM). The optical transmittance of the suspended flake increases from 0.4 to 0.9 by the laser exposure, indicative of a reduced absorption coefficient of the material as expected for tantalum oxide.

To investigate the changes in chemical composition caused by laser-irradiation of $TaSe_2$ we carried out Elemental Dispersive X-ray (EDX) measurements on freely suspended areas of the material. Composition analysis shows an approximate ratio of Ta/Se/O=1.0/2.3/0.4 in the pristine flake. The presence of oxygen in EDX is attributed to surface oxidation of the $TaSe_2$ under atmospheric condition (see Section V in the ESM) [12]. Interestingly, after laser exposure the composition of the flake is Ta/Se/O=1.0/0.0/3.8, which indicates that the laser-irradiation procedure in air oxidizes the flake, removing the selenium atoms and replacing them by oxygen. Thus, laser exposure converts the $TaSe_2$ into tantalum oxide. Note, that this compositional analysis has an uncertainty of ±20% hampering the determination of the exact stoichiometry after laser-oxidation. So even though the O/Ta ratio determined by EDX is too large for $Ta_2O_5$, one cannot rule out that this tantalum oxide is $Ta_2O_5$ due to uncertainties in EDX and the presence of surface oxides.

In order to determine the effect of laser oxidation on the flake's crystal structure, we performed a TEM analysis on the suspended flake. Fig. 1(b) shows a bright field TEM image of the region highlighted with a square in Fig. 1(a). The transition boundary between the pristine (top) and oxidized (bottom) part of the flake is abrupt (<50 nm). A High Resolution TEM (HRTEM) image of the pristine suspended $TaSe_2$ is shown in Fig. 1(c), demonstrating a hexagonal atomic configuration, which is confirmed by Selective Area Diffraction Pattern (SADP) analysis (see Section V in the ESM). The crystal structure and lattice constant from HRTEM and SADP are consistent with an in-plane orientation of the layered $\{1\bar{1}00\}$ planes of 2H-$TaSe_2$ with an interatomic Ta-Ta spacing of 3.44 Å (inset). Figure 1(d) presents a HRTEM image of the oxidized region of the suspended material. Both amorphous and crystalline regions are observed. In the crystalline domains, HRTEM (inset) and SADP show a hexagonal configuration with an interatomic Ta-Ta spacing of 3.63 Å, larger than in $TaSe_2$. The hexagonal structure and Ta-Ta distance obtained from TEM of laser oxidized $TaSe_2$ matches that of TT-$Ta_2O_5$ (also called δ-$Ta_2O_5$) [13–15]. Furthermore, the Raman spectrum and photoluminescence spectra (see Section IV in the ESM) of the oxidized flake correspond to that of $Ta_2O_5$ [16]. These observations suggest that the modification is the oxidation reported in $TaSe_2$ when heated up to around $T_{ox}$=600 °C [17], which results in a crystal structure that bears most resemblance to TT-$Ta_2O_5$.

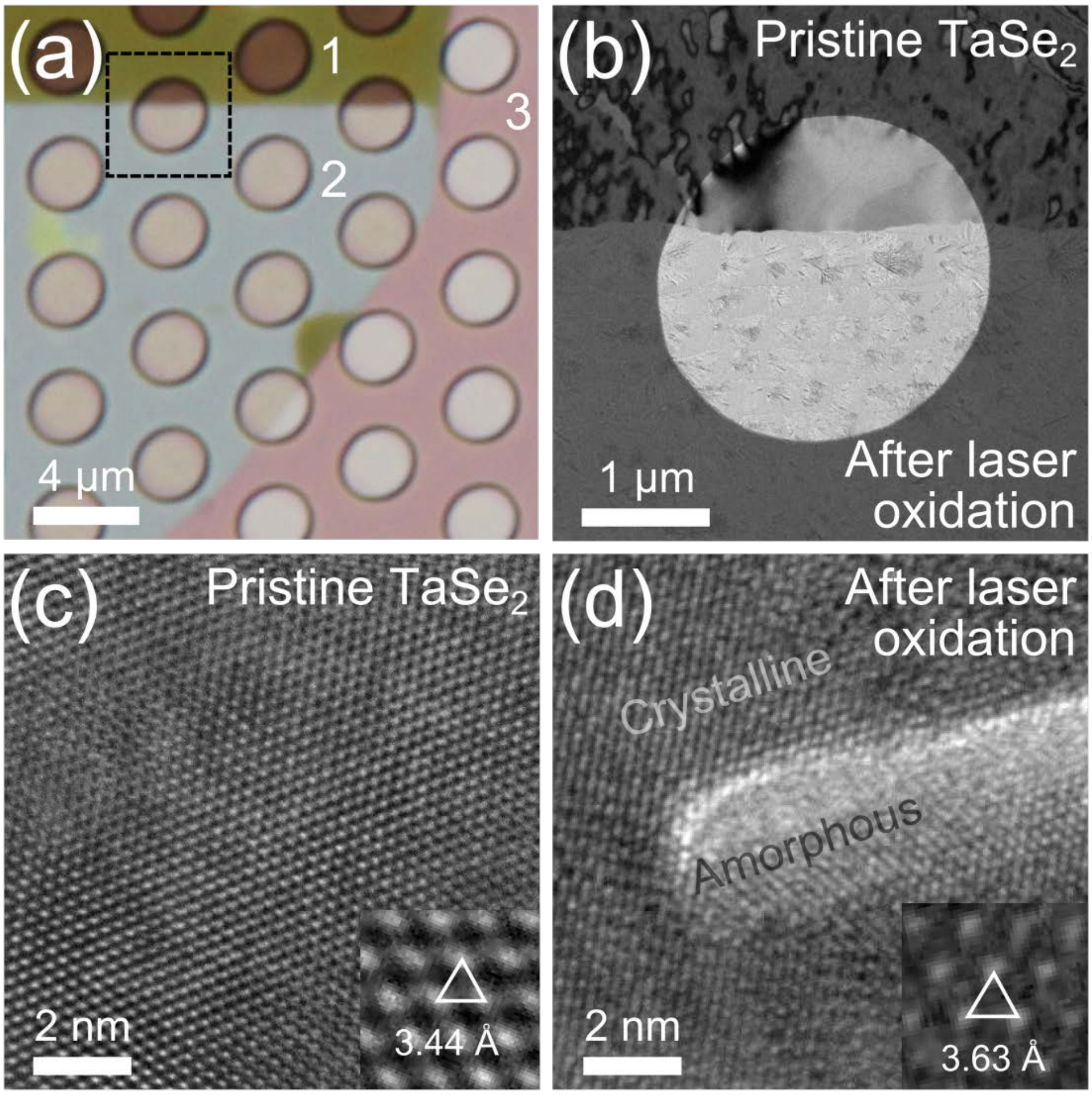

**Figure 1** Characterization of a $TaSe_2$ flake before and after laser-oxidation. (a) Transmission optical image showing pristine (1) and laser-oxidized (2) regions of the $TaSe_2$ flake on a silicon nitride membrane TEM grid (3) with circular holes. (b) Bright field TEM image of the partially oxidized suspended flake indicated by the dashed box in (a). (c) HRTEM image of suspended pristine $TaSe_2$. The interatomic distance (3.44 Å) corresponds to the Ta-Ta distance in the $\{1\underline{1}00\}$ plane of 2H-$TaSe_2$ [12]. (d) HRTEM image of a laser-oxidized region showing coexisting amorphous and crystalline phases. The interatomic distance in the hexagonal crystal plane has increased to 3.63 Å.

The oxidation process in the suspended $TaSe_2$ happens as follows. Once the temperature of $TaSe_2$ is increased over that critical temperature $T_{ox}$ by laser-heating, it oxidizes and becomes more transparent. The increased transparency reduces light-absorption and thus leads to a temperature reduction in the flake. Thus, despite inhomogeneities in the laser power and thermal resistances over the flake, this self-limiting mechanism prevents the flake from exposure to temperatures significantly higher than $T_{ox}$. This effect also prevents ablation [18], and improves the homogeneity of both the composition and stress in the oxidized film.

After having established the effects of laser-oxidation on the composition and crystal structure of $TaSe_2$, we now study its effects on the mechanical properties of drumhead resonators. $TaSe_2$ drumhead resonators are fabricated by transferring mechanically exfoliated $TaSe_2$ flakes onto a $SiO_2$/Si substrate with circular cavities of 3.2 μm in diameter and 285 nm in depth. The mechanical properties of the drumhead resonators are investigated by measurement of their fundamental mechanical resonance mode using an optical interferometer setup [9, 10] (see Section VI in the ESM). The mechanical spectrum is measured over a wide frequency range as shown in Section IX, although we use only the fundamental mode for the analyses.

Figure 2 shows the mechanical resonance spectrum of a 17 nm thick $TaSe_2$ drumhead resonator vibrating in its fundamental mode before and after laser-oxidation. The insets show the corresponding optical images of the device, and the square-shaped laser-exposed area containing the suspended drumhead in the oxidized case. The resonance frequency increases from $f_{pris}$=10.2 MHz to $f_{ox}$=39.6 MHz after oxidation, where $f_{pris}$ and $f_{ox}$ are the fundamental frequencies for the pristine and oxidized cases, respectively. At the same time, the Q factor rises from $Q_{pris}$=357 to $Q_{ox}$=1058. A driven harmonic oscillator model fits the data and is used to determine the Q factor (solid lines). The magnitude of the interferometer signal after oxidation is a factor 3700 smaller than the signal before oxidation. This is attributed to a reduction of the photothermal actuation efficiency by a diminished optical absorption and to a reduction of the interferometric signal by reduction of the drum's reflectance after oxidation. After having demonstrated a controlled enhancement of $f_0$ and $Q$, we proceed to perform a systematic study on devices of different thickness.

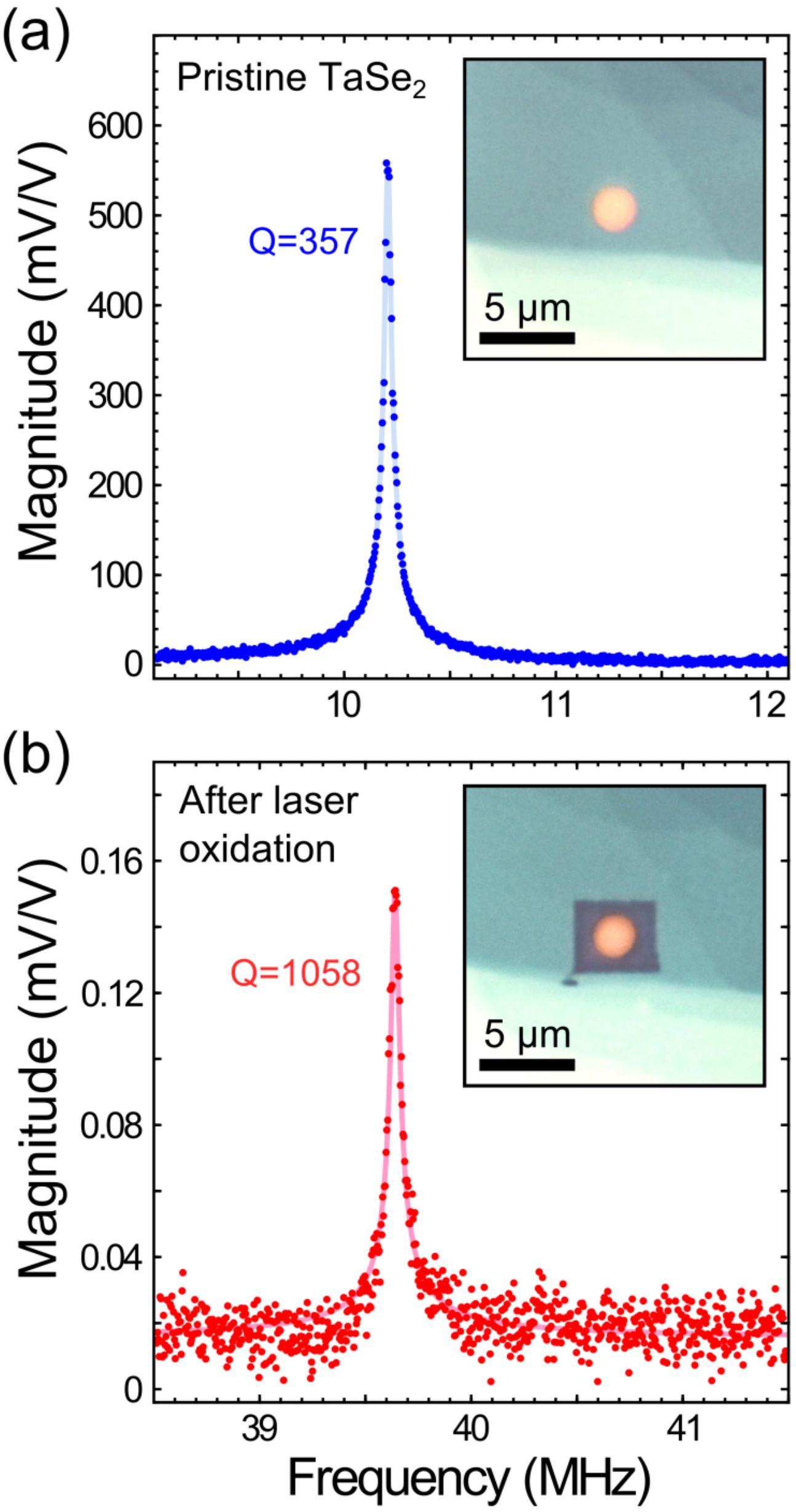

**Figure 2** Mechanical resonances of $TaSe_2$ flakes before and after laser-oxidation. (a) Fundamental mechanical resonance peak of a pristine 3.2 μm diameter drum with a thickness of 17 nm. Inset: Optical image of the drum. (b) Fundamental resonance peak of the same drum after laser-oxidation of the square region shown in the inset. A large enhancement of the resonance frequency and Q factor is observed.

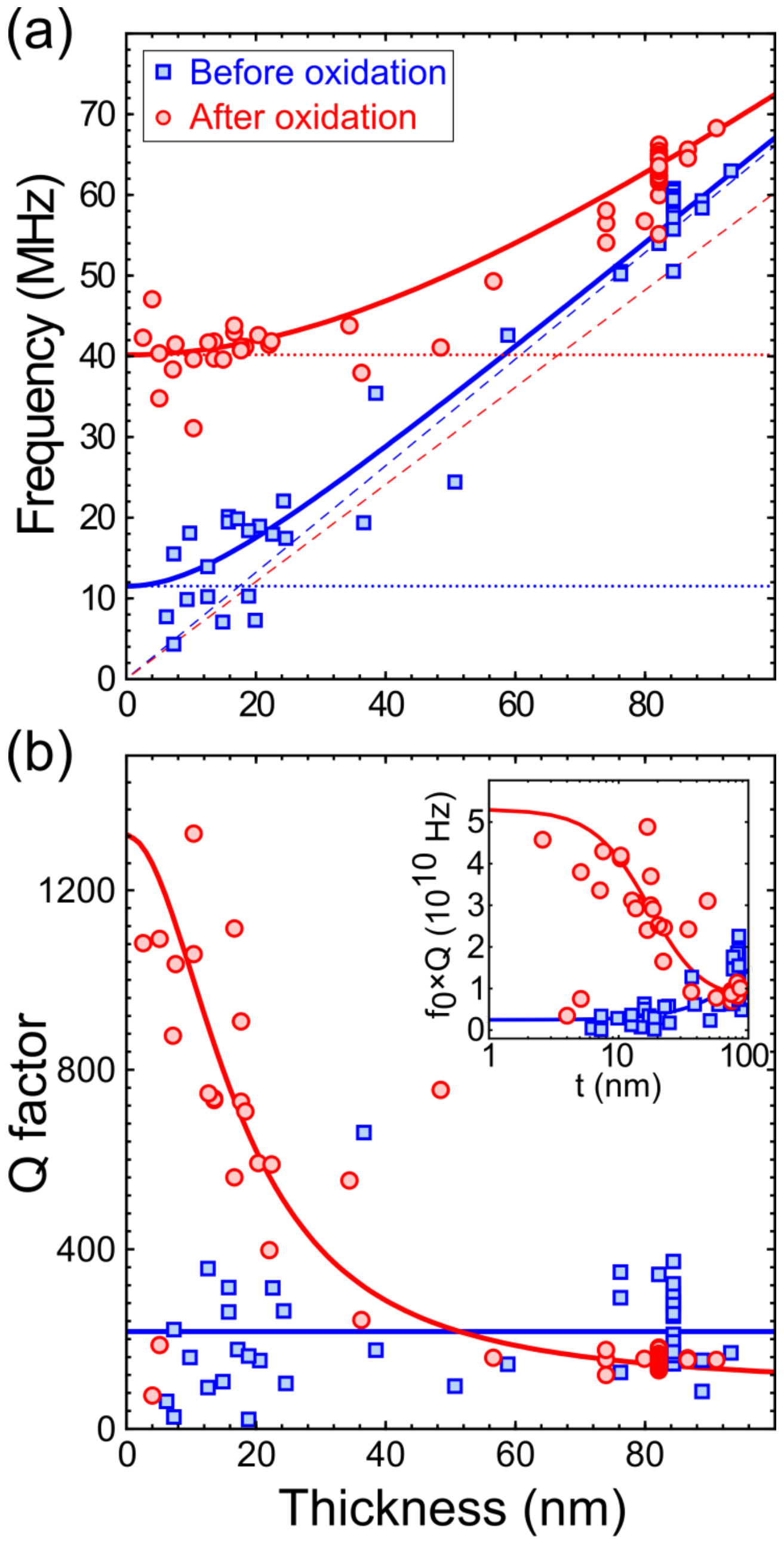

**Figure 3** Thickness dependence of $f_0$ and $Q$ of several drum resonators before and after laser-oxidation. (a) Resonance frequency versus thickness of drums with a diameter of 3.2 μm from the same $TaSe_2$ flake before (blue squares) and after oxidation (red circles). Equation 2 (solid lines) is used to fit the data. For small thickness the resonance frequency follows the membrane limit (horizontal dotted lines, Eq. 1), whereas for thick drums the data converges toward the plate limit (dashed lines, Eq. 1). (b) Q factor versus thickness: for thin drums an increase in $Q$ is observed after laser-oxidation. Solid red lines correspond to fits to the data by Eq. 3. Inset: In thin drums, laser-oxidation increases the $f$-$Q$ product.

Figure 3 shows the resonance frequencies and Q factors of 3.2μm -in-diameter drums with thicknesses ranging from 6 to 89 nm, all from the same flake to ensure that they have the same built-in pre-stress, before (blue squares) and after laser-oxidation (red circles). Starting with the pristine drums (blue data), we observe for thick devices a linear relation between resonance frequency and thickness because the bending rigidity dominates the mechanics of the resonator (dashed lines represent this plate-like mechanical behaviour). In the limit of small thickness, the resonance frequency converges to a constant value because the pre-stress dominates the mechanics (dotted lines represent this membrane limit). In the oxidized drums (red circles in Fig. 3(a)), the resonance frequency behavior in the plate limit is similar to that of the pristine devices. However, the thin oxidized drums show a much higher resonance frequency in the membrane limit, which indicates a larger pre-stress. The complete dataset is presented in a table in Section VII from the ESM.

To extract the stress increase caused by laser-oxidation, we modeled the resonance frequency of the drums as follows. For thin drums, the fundamental resonance frequency of the drums converges to the membrane limit $f_{\text{mem}}$, whereas for thick drums the frequency converges to the plate limit $f_{\text{plate}}$ [19]:

$$f_{\text{mem}} = \frac{2.40}{\pi d}\sqrt{\frac{\sigma}{\rho}} \qquad f_{\text{plate}} = \frac{10.21}{\pi}\sqrt{\frac{E}{3\,\rho(1-\nu^2)}}\frac{t}{d^2}, \qquad (1)$$

where $E$ is Young's modulus, $\nu$ is Poisson's ratio ($\nu \approx 0.2$) [20], $\rho$ is the mass density ($\rho_{\text{pris}}$ = 8660 kg m$^{-3}$, $\rho_{\text{ox}}$= 0.655 $\rho_{\text{pris}}$ [17]), $t$ is the thickness, $d$ is the resonator diameter, and $\sigma$ is the pre-stress in the drumhead. For drums of intermediate thickness, the resonance frequency can be approximated by addition of the spring constants of plate and membrane modes giving:

$$f_0 \approx \sqrt{f_{\text{mem}}^2 + f_{\text{plate}}^2}\,. \qquad (2)$$

A fit of Eq. 2 to the data in Fig. 3(a) (solid lines) gives an estimate of the Young's modulus ($E_{\text{pris}}$=110 GPa and $E_{\text{ox}}$=60 GPa) and of the stress ($\sigma_{\text{pris}}$=20 MPa and $\sigma_{\text{ox}}$=160 MPa) values before and after oxidation. The estimated Young's modulus for $TaSe_2$ is in good agreement with values found in literature [21–26]. Since the crystal structure of the oxidized flake consists of a mixture of amorphous and crystalline regions it is not possible to make a comparison between the Young's modulus of the oxidized flake and the data available in the literature. The fit shows a large increase of the tensile stress in the membrane from 20 MPa to 160 MPa. The drastic increase in the resonance frequency during oxidation of the thin $TaSe_2$ drums can be mainly attributed to the increase in tensile stress.

Having established clearly the presence of oxidization-induced stress, we now address a possible mechanism for how this stress is created. The laser-oxidation occurs locally and only heats up the material without significantly affecting the substrate beneath. During the recrystallization of the oxide at high temperature $T_{\text{ox}}$=600 °C, atoms rearrange, which leads to stress relaxation in the suspended part of the drum. While cooling down, thermal contraction of the drum increases the stress in the membrane. An estimated value of the resulting tensile stress, using the coefficient of thermal expansion $\alpha_{\text{ox}}$=3.6x10$^{-6}$ of $Ta_2O_5$ [27] yields

$\sigma_{\mathrm{ox,e}} \approx (T_{\mathrm{ox}} - 25\ °C)\alpha_{\mathrm{ox}}E_{\mathrm{ox}} = 120$ MPa, in agreement with the measured stress $\sigma_{ox}$=160 MPa.

In addition to changing the frequency of the mechanical resonators, the oxidization process also significantly changes the mechanical quality factor. Figure 3(b) shows the thickness dependence of the quality factor of the fundamental resonance before and after laser-oxidation. We find that the quality factor in pristine $TaSe_2$ resonators (blue squares) is almost independent of the thickness. For laser-oxidized drums on the other hand, the quality factor shows a strong thickness dependence: in thin oxidized drums the quality factor is up to a factor 9.4 higher than that in thick oxidized drums.

In the following, we show that the increase of the Q factor originates from the large tensile stress of the thin oxidized drums. A similar increase in Q factor in highly stressed thin membranes has been observed in $Si_3N_4$ resonators and reduced graphene oxide [2, 6, 7, 28–31]. The almost thickness- independent Q factor in low-stress (plate-like) drums observed in the pristine material can be phenomenologically described [6, 7] using a material model with a frequency-independent complex Young's modulus $E = E_1 + iE_2$. The imaginary part of the Young's modulus causes dissipation. Since the bending rigidity is proportional to the Young's modulus, the complex plate spring constant will be given by $k_{\mathrm{plate}} = k_{\mathrm{plate,1}} + \mathrm{i}k_{\mathrm{plate,2}}$. The resonator's Q-factor in the plate limit will therefore be given by $Q_{\mathrm{plate}} = \frac{E_1}{E_2} = \frac{k_{\mathrm{plate,1}}}{k_{\mathrm{plate,2}}}$. Similarly, in the membrane limit the Q factor can be expressed as $Q_{\mathrm{mem}} = \frac{k_{\mathrm{mem,1}}}{k_{\mathrm{mem,2}}}$, however since the losses due to elongation are much lower than the losses due to bending we will assume that $Q_{\mathrm{mem}} \gg Q_{\mathrm{plate}}$ and $k_{\mathrm{mem,2}} \approx 0$. By adding the bending and membrane spring constants (as in Eq. 2), the stress dependence of the Q factor is therefore given by $= \frac{k_{\mathrm{mem,1}} + k_{\mathrm{plate,1}}}{k_{\mathrm{plate,2}}}$. This relation shows that a stress-induced increase of $k_{mem,1}$, will result in an increase of the resonator's Q factor.

Based on the relation $Q = \frac{k_{\mathrm{mem,1}} + k_{\mathrm{plate,1}}}{k_{\mathrm{plate,2}}}$ an equation for the $Q$-ratio between pristine and oxidized drums is derived in Section VIII from the ESM:

$$\frac{Q_{\mathrm{ox}}}{Q_{\mathrm{pris}}} = \alpha \left(\frac{f_{\mathrm{ox}}}{f_{\mathrm{pris}}}\right)^2, \qquad (3)$$

where the coefficient $\alpha$ is defined as $\alpha = \frac{m_{\mathrm{ox}}}{m_{\mathrm{pris}}} \frac{E_{2,\mathrm{pris}}}{E_{2,\mathrm{ox}}}$ and $m$ is the mass of the drumhead. Equation 3 is used to fit the thickness-dependence of $Q_{ox}$ in Fig. 3(b) (solid lines). The values of $f_{pris}$ and $f_{ox}$ determined from the fits in Fig. 3(a) and an average value $Q_{pris}$=216 are used. By adjusting the coefficient $\alpha = 0.5$ as the only fit parameter, the thickness dependence of $Q_{ox}$ is well captured by Eq. 3. The model based on a frequency-independent complex Young's modulus, that was proposed to model the Q-factor increase with stress in $Si_3N_4$ beams [1, 7, 30, 32], is thus observed to be consistent with the thickness dependence of the Q-factor in oxidized $TaSe_2$ flakes (Eq. 3).

**Conclusions**

The *f-Q* product is an important figure of merit for micro and nanoresonators, because high *f-Q* products can yield low phase-noise high frequency oscillators and sensors. Since tensile stress has been shown to increase both resonance frequency and Q factor, the presented laser oxidation procedure is a very effective method to increase the *f-Q* product as is shown in the inset of Fig. 3(b). The increase is as high as a factor 42 yielding a maximum *f-Q* product of $4.9\times10^{10}$ Hz. To our knowledge the presented laser-oxidation method yields the highest *f-Q* product at room temperature in ultrathin resonators ($t$<20 nm) made out of 2D materials, outperforming *f-Q* products reported in graphene and $MoS_2$ devices at room temperature [33].

In summary, a laser-oxidation procedure for enhancement of quality factor and resonance frequency of multilayer $TaSe_2$ resonators is demonstrated. The procedure increases the stress in the drums by a factor 8, due to thermal contraction during cooldown after laser-oxidation. The stress results in an enhanced resonance frequency (up to 9 times larger) and Q factor (over 14 times larger), which is attributed to a stress-induced increase in spring constant. The presented laser oxidation procedure thus provides a tool to locally and selectively modify the mechanical properties of 2D materials. This enables interesting applications, such as in-situ tuning of the resonance frequency and the Q factor, and, by the selective patterning of stress regions in suspended flakes, engineering the mechanical mode shapes and intermode relations (see Section IX and X in the ESM). The same procedure is expected to be applicable in other 2D materials, as many metal dichalcogenides show oxidation reactions similar to $TaSe_2$ [17, 34].

## Acknowledgements

The research leading to these results has received funding from the European Union Seventh Framework Programme under grant agreement no 604391 Graphene Flagship. A.C.-G. acknowledges financial support through the FP7-Marie Curie Projects PIEF-GA-2011-300802 ('STRENGTHNANO').

**Electronic Supplementary Material**

Supplementary material is available in the online version of this article at http://dx.doi.org/10.1007/s12274-***-****-*.

Electronic Supplementary Material

# High-Q Tantalum Oxide Nanomechanical Resonators by Laser-Oxidation of $TaSe_2$

Santiago J. Cartamil-Bueno[1] (✉), Peter G. Steeneken[1] (✉), Frans D. Tichelaar[2], Efren Navarro-Moratalla[3], Warner J. Venstra[1], Ronald van Leeuwen[1], Eugenio Coronado[3], Herre S.J. van der Zant[1], Gary A. Steele[1], and Andres Castellanos-Gomez[1,†] (✉)

[1] Kavli Institute of Nanoscience, Delft University of Technology, Lorentzweg 1, 2628 CJ Delft, The Netherlands

[2] Delft University of Technology, Kavli Institute of Nanoscience, National Centre for HREM, Lorentzweg 1, 2628 CJ Delft, The Netherlands

[3] Instituto Ciencia Molecular (ICMol), Univ. Valencia, C/Catedrático José Beltrán 2, E-46980, Paterna, Spain

[†] Instituto Madrileño de Estudios Avanzados en Nanociencia (IMDEA-Nanociencia), 28049 Madrid, Spain

Synthesization of single-crystal $TaSe_2$

$TaSe_2$ crystals are synthesized from the elemental components in a two-step process. Polycrystalline $TaSe_2$ is obtained by ceramic combination of stoichiometric ratios of Ta and Se. Ta powder, 99.99% trace metals basis and Se powder, -100 mesh, 99.99% trace metals basis are used. Powdered starting materials are intimately mixed, placed inside an evacuated quartz ampoule and reacted at 900 °C for 9 days. The resulting free-flowing glittery grey microcrystals are then transformed into large single-crystals using the chemical vapor transport (CVT) methodology. For that purpose, 1 g of $TaSe_2$ polycrystalline material together with 275 mg of $I_2$ are loaded into a 500 mm long quartz ampoule (OD: 18 mm, wall-thickness: 1.5 mm). The mixture is placed at one end of the ampoule which is exhaustively evacuated and flame-sealed. The quartz tube is finally placed inside a three-zone split muffle where a gradient of 25 °C is established between the leftmost load (725 °C) and central growth (700 °C) zones. A gradient of 25 °C is also set between the rightmost and central regions. The temperature gradient is maintained constant during 15 days and the muffle is eventually switched off and left to cool down to ambient conditions. Millimeter-size $TaSe_2$ crystals are recovered from the ampoule's central zone, exhaustively rinsed with diethyl ether and stored under a $N_2$ atmosphere.

Fabrication and laser-oxidation of $TaSe_2$ resonators

Freely suspended $TaSe_2$ layers are fabricated by mechanical exfoliation of $TaSe_2$ crystals onto a substrate with microcavities. The substrate is fabricated from Si wafers with 285 nm of thermally grown $SiO_2$, and contains circular cavities of that depth and a diameter of 3.2 micron. The $SiO_2$ thickness of 285 nm is ideal as the optical contrast of the thinner flakes is enhanced.

As recently demonstrated for $MoS_2$ flakes [S1], the all-dry transfer technique based on elastomeric stamps was employed. The process is summarized as follows:

$TaSe_2$ flakes are deposited on an elastomeric stamp (GelFilm® by GelPak) by mechanical exfoliation of synthetic $TaSe_2$ with blue Nitto tape (Nitto Denko Co., SPV 224P).

$TaSe_2$ flakes of different thicknesses are identified on the surface of the viscoelastic stamp by transmission mode optical microscopy.

The stamp surface containing the $TaSe_2$ flakes is mounted in a micromanipulator facing the patterned substrate.

The stamp is brought into contact with the patterned substrate by lowering the manipulator.

The stamp is peeled-off very slowly using the micromanipulator.

The resulting resonators are inspected with an optical microscope (Olympus BX 51 supplemented with a Canon EOS 600D camera) as shown in Figure S-1.

A Renishaw in via system is used to scan a focused laser spot ($\lambda$ = 514 nm) over the $TaSe_2$ drums, resulting in square-shaped oxidized regions, as displayed in the right part of Figure S1. The oxidation of the material occurs at a laser power of 25 mW for 0.1-0.2 s of exposure time.

Atomic force microscopy characterization

A scanning probe microscope (Digital Instruments D3100 AFM) with cantilevers (BudgetSensors, spring constant 40 N $m^{-1}$, tip curvature <10 nm) is operated in the amplitude modulation mode to study the topography and to determine the thickness of the flakes. The thickness of the drums is determined both before and after laser-oxidation by measuring step heights. The measured thickness reduction after laser-oxidation is typically 3±1 nm, independent of the initial flake thickness.

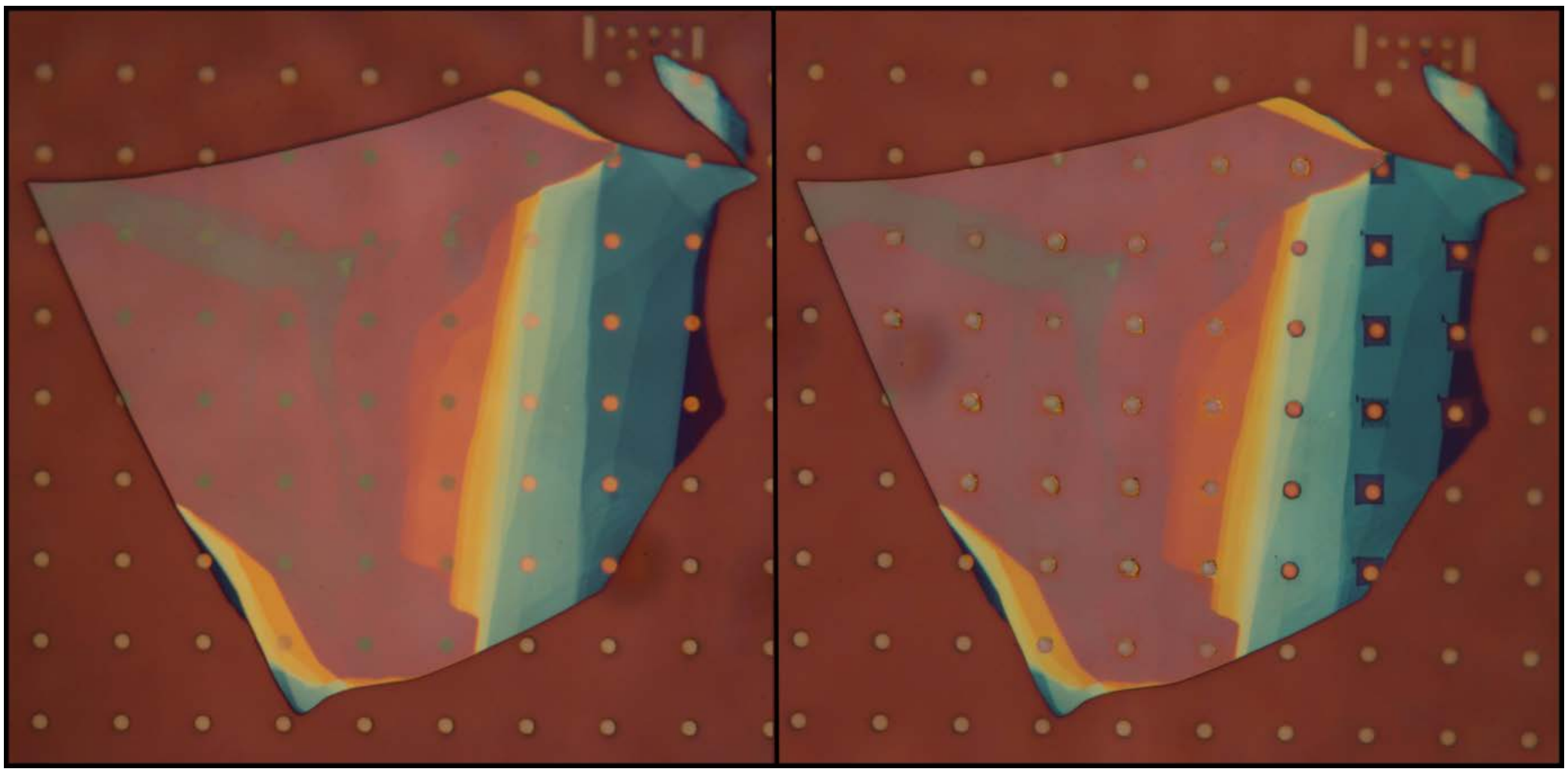

**Figure S-1.** Optical images of the $TaSe_2$ flake on a Si substrate with circular cavities before (left) and after (right) oxidation.

Raman and photoluminescence characterization

Raman spectroscopy on a $TaSe_2$ flake over $SiO_2$/Si was carried out with the same Renishaw in Via Raman microscope that was used for the laser oxidation in order to obtain more information about the material after laser exposure. Measurements were performed in backscattering configuration excited with visible laser light ($\lambda$ = 514 nm) on the suspended part of a $TaSe_2$ drum and on its surroundings [S3]. The spectra are collected through a 100× objective (NA = 0.95) and recorded with a 1800 lines $mm^{-1}$ grating providing a spectral resolution of ~0.5 $cm^{-1}$. During the characterization, the laser power is kept at low power levels, P ~ 250 μW. Raman spectroscopy revealed that the suspended part was more easily oxidized and with less exposure time than the regions on top of the substrate. This is attributed to a higher thermal resistance of the suspended part of the flake, as a consequence of which less heating power is needed to reach the oxidation temperature. Figure S-2 displays the Raman spectrum for $TaSe_2$ before laser-exposure in a suspended region. The Raman spectrum shows the $E_{12g}$ and $A_{1g}$ peaks and a background signal. After laser-exposure (Figure S-3), a feature around 255 $cm^{-1}$ appears. This Raman peak is at the same energy as the most characteristic Raman peak for crystalline $Ta_2O_5$ in that part of the spectrum [S2]. Moreover, two broad peaks around 1350 $cm^{-1}$ and 1550 $cm^{-}$ 1 develop, corresponding to a photoluminescence emission around 550 nm to 560 nm, respectively. Similar photoluminescence emission has been reported for $Ta_2O_5$ [S4, S5].

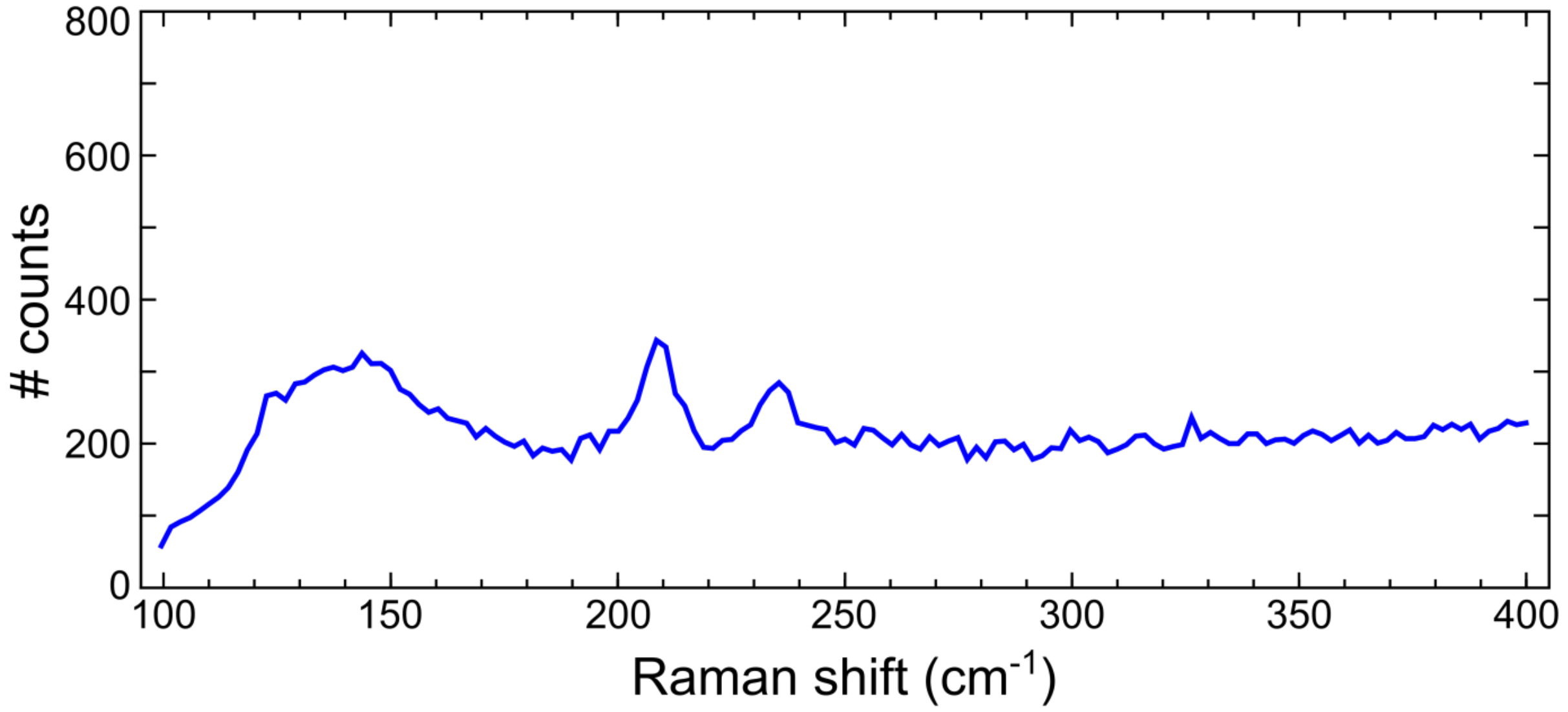

**Figure S-2.** Raman spectrum of a $TaSe_2$ flake in a suspended region.

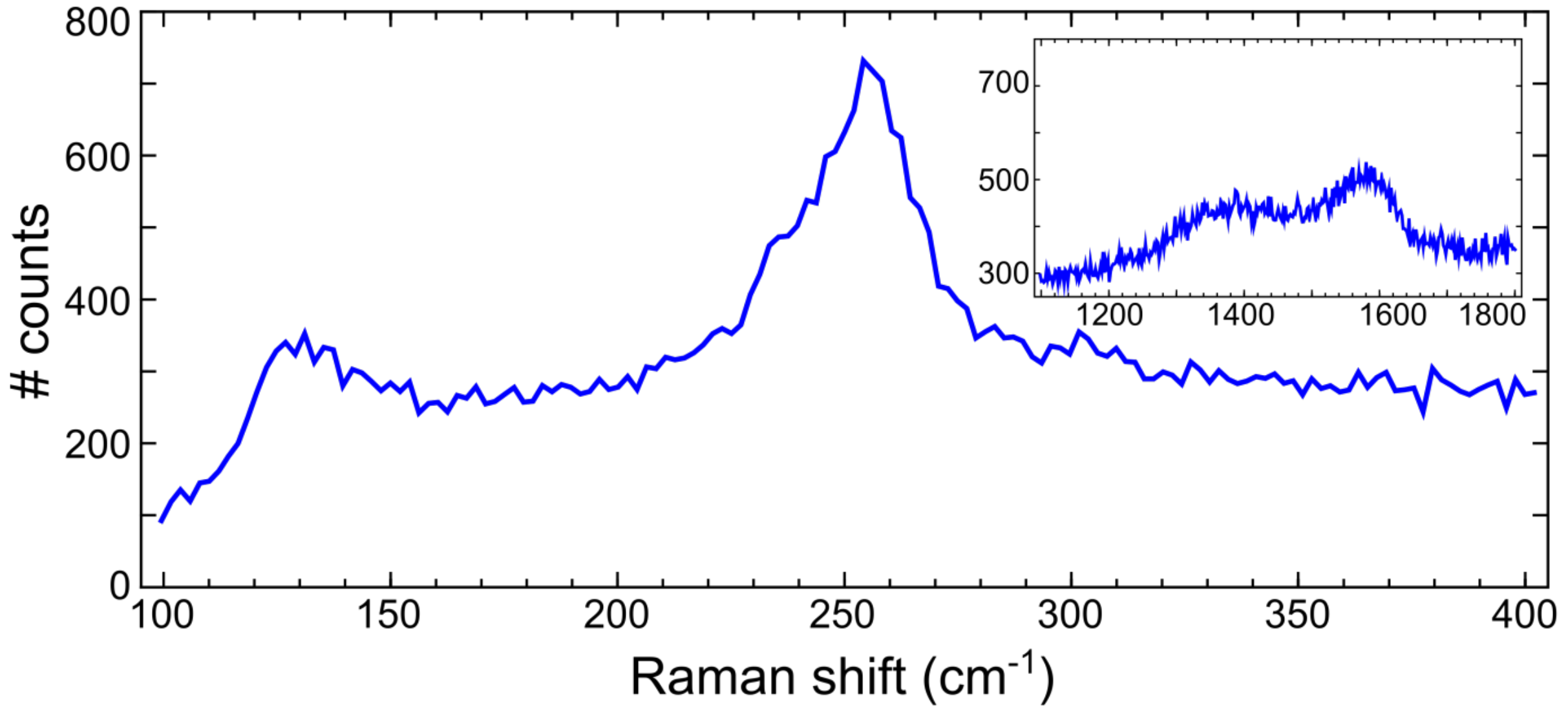

**Figure S-3.** Raman spectrum of a laser-oxidized $TaSe_2$ region.

Transmission Electron Microscopy (TEM) characterization

A Tecnai F20ST/STEM (200 kV) was used for TEM imaging. Bright-field TEM images of the laser-treated and pristine areas of a $TaSe_2$ flake are shown in Figure 1b. The $Si_3N_4$ TEM grid is visible in Figure 1a-b.

*Selected Area Diffraction Pattern (SADP) analysis*

Figure S-4a shows SADP patterns of the crystalline pristine and oxidized flake. The lattice spacing of the crystals can be deduced according to $d=\lambda L/D$, where $d$ is the lattice spacing, $\lambda$ is the electron wavelength, $L$ the camera length and D the distance measured on the CCD detector from the reflected beam to the undiffracted beam. The measurement uncertainty is usually 1-2%. The lattice spacing between the crycstal planes in the pristine flake is found to be $d_{pris}$=2.98 Å, in correspondence with a Ta-Ta spacing of $\frac{2d}{\sqrt{3}} = 3.44$ Å in 2H-TaSe [S6].

After oxidation, the SADP becomes more complex, as is shown in Figure S4b. The dominant hexagonal pattern shows $d$=3.14 Å as lattice spacing, rather different from the pristine part (2.98 Å). In between the dominant reflections vague rings exists, indicative of small ordered areas. Ordered because they correspond to twice the 3.14 Å spacing, and small because the reflections are wide. In the background, vague rings can be seen, that correspond to amorphous regions as is confirmed from a separate SADP on the amorphous area showing similar amorphous rings.

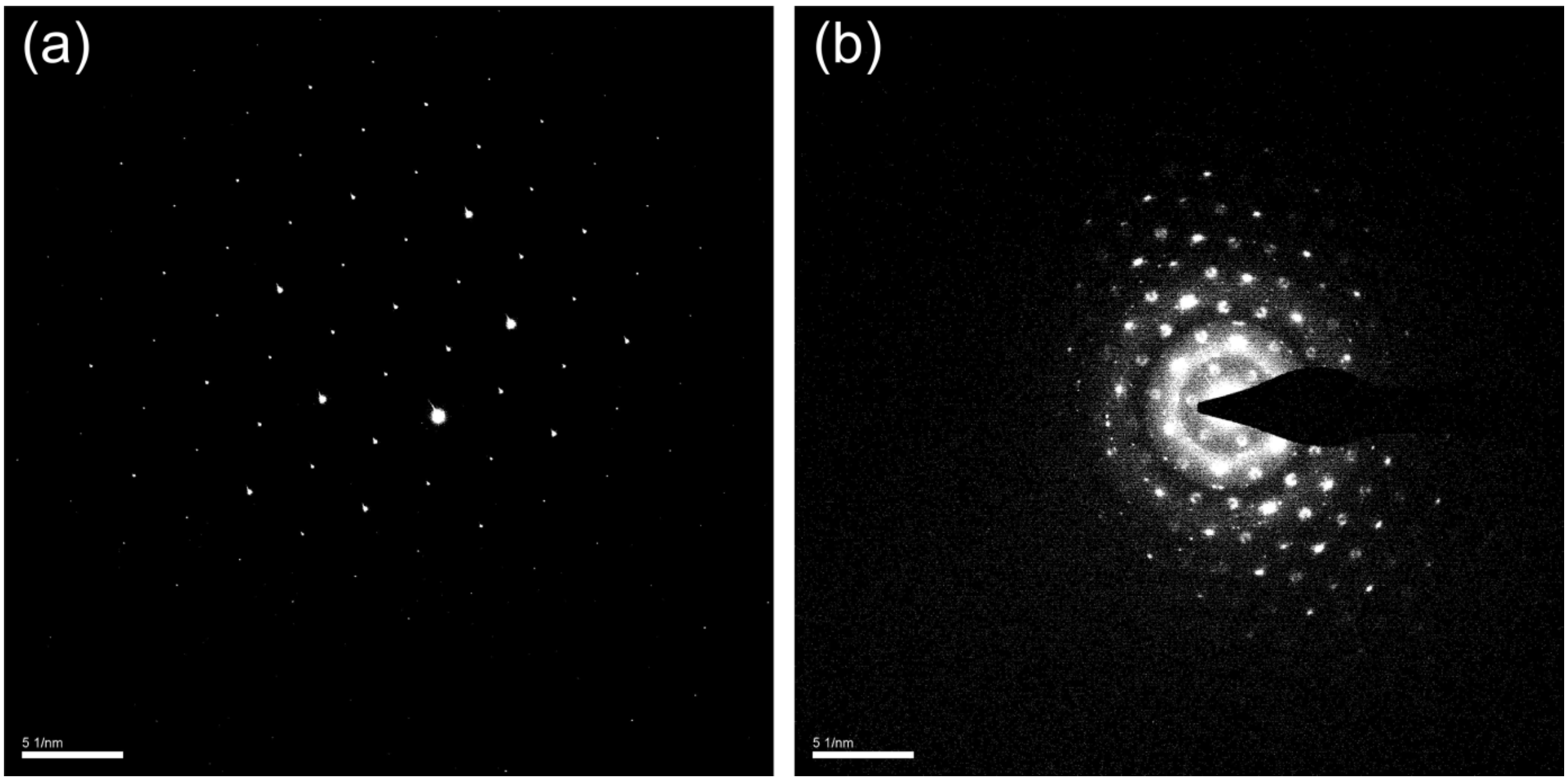

**Figure S-4.** (a) SADP of the pristine single crystal flake. The reflections close to the origin are of type <1$\bar{1}$00> with d spacing 2.98 Å. (b) SADP of a laser-oxidized area.

*Energy-dispersive X-ray spectroscopy (EDX) characterization*

EDX data taken on the pristine flake and on both crystalline and amorphous regions of the oxidized flake are shown in Table S-5.

| | O | Se | Ta |
|---|---|---|---|
| Pristine 1 | 62 | 24 | 11 |
| Pristine 1 | 12 | 61 | 27 |
| Oxide crystalline 1 | 80 | 1.7 | 19 |
| Oxide crystalline 2 | 82 | 0.3 | 17 |
| Oxide crystalline 3 | 77 | 0.2 | 23 |
| Oxide amorphous 1 | 75 | 1.7 | 23 |
| Oxide amorphous 2 | 75 | 0.7 | 24 |

**Table S-5.** EDX atomic concentration quantification before and after laser oxidation. Atomic % relative uncertainty is 20%.

*HRTEM characterization*

Figure S-6 shows a HRTEM picture shows a region in the suspended flake containing both amorphous and crystalline regions. The brighter amorphous regions form dendritic structures. Figure S-7 shows an HRTEM image of pristine and oxidized regions of suspended edges of the flake.

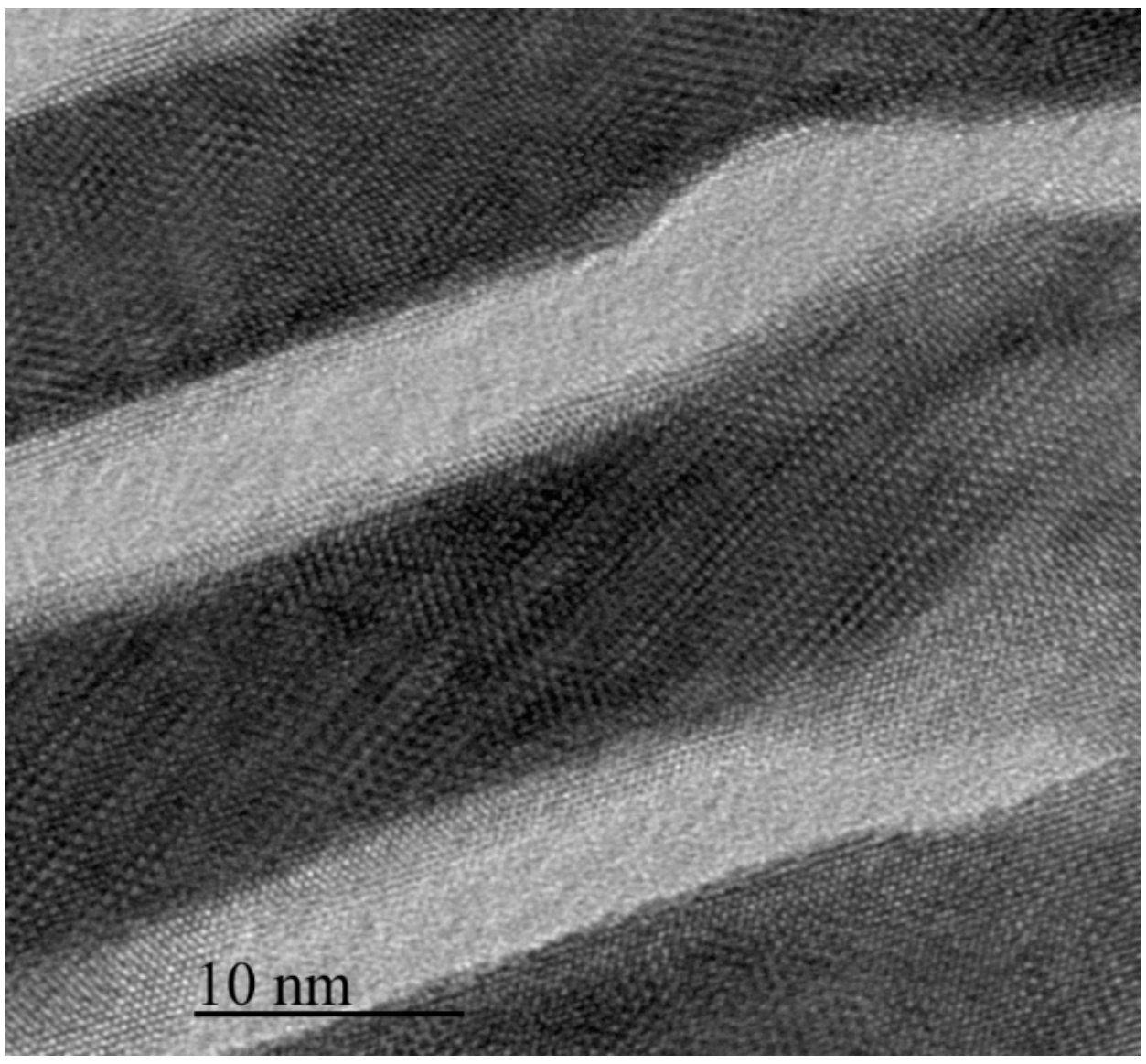

**Figure S-6.** HRTEM image of dendritic structure showing both crystalline and amorphous regions.

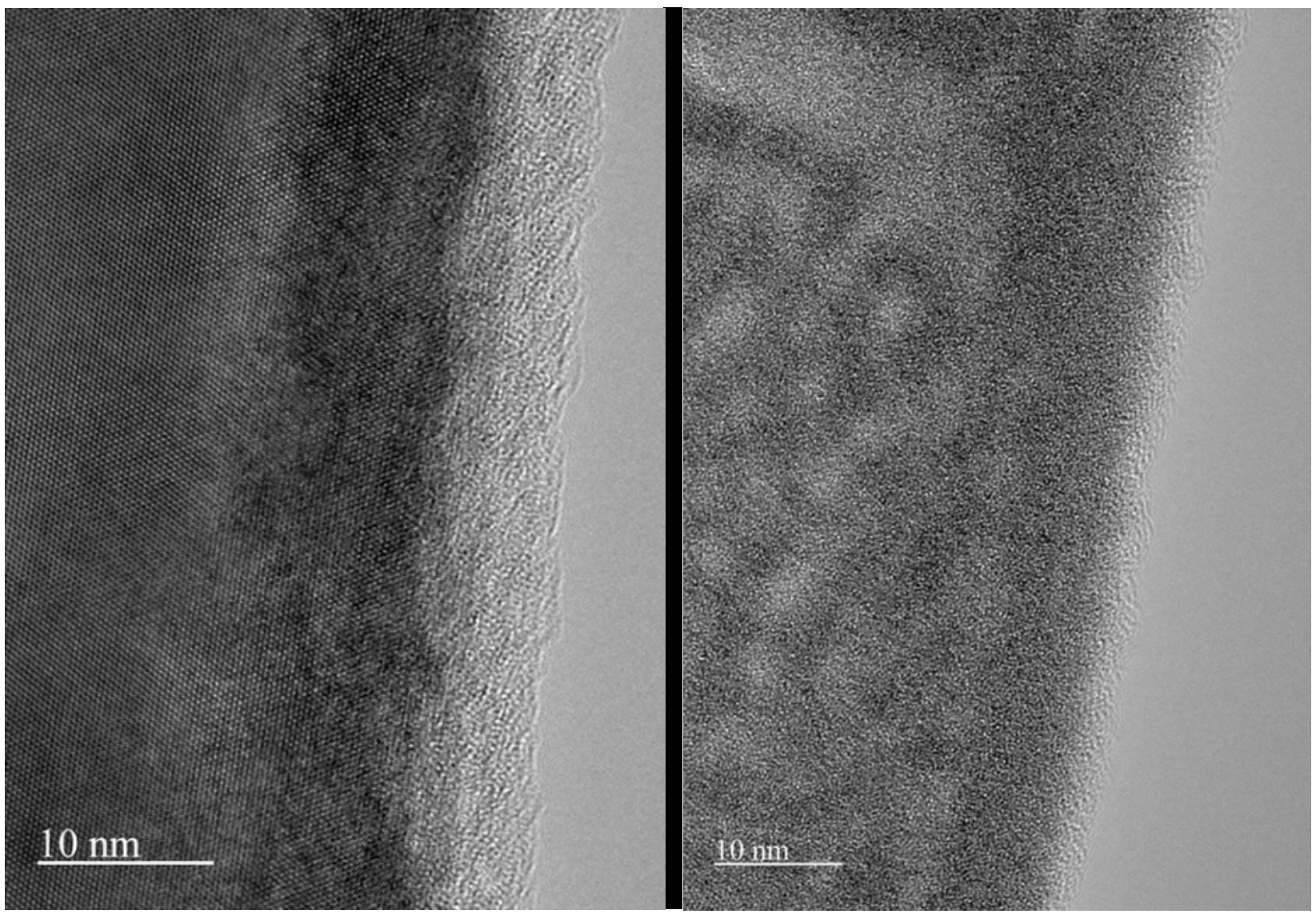

**Figure S-7.** HREM images of suspended edges of the pristine (left) and oxidized (right) flake

<u>Optical interferometer setup</u>

The frequency response of the resonators is measured using an optical interferometer [S7, S8]. The suspended material is photothermally actuated with an intensity modulated blue diode laser (Thorlabs LP405-SF10, $\lambda$ = 405 nm) at a power of 260 μW (CW), while a second constant-power linearly-polarized Helium-Neon laser ($\lambda$ = 632.8 nm) at 4 mW is used as optical probe for the detection. The measurements are carried out in vacuum (~$10^{-5}$ mbar) at room temperature.

The power from the He-Ne laser is adjusted with a neutral density filter, and a beam expander enlarges the beam diameter by a factor of three to match the aperture of the objective lens. The beam passes a polarized beam splitter and a quarter wave plate, and is combined with.

a blue laser, which is intensity-modulated (from a typical optical modulation depth of 4 μW as measured at the diode laser output, up to 137 μW for the oxidized drums) for high frequency photothermal excitation. The combined red and blue light is focused on the suspended drumhead with a microscope objective lens (Leica 50×, NA=0.6).

The reflected light goes back through the objective lens, and part of it is sent to a CCD camera (due to the beam splitter 50/50) which is used for visual inspection and alignment. The reflected blue beam is taken out of the main path by the dichroic mirror. The reflected red light is phase-shifted again and directed towards a high speed photo-detector (NewFocus 1601) as explained above. The detector output is connected to the input port of a network analyzer (Rohde & Schwartz, ZVB4).

As the suspended material is very thin, most of the light is transmitted through the flake and reflected at the bottom of the cavity. However, a small part of the light is reflected at the material surface, and it interferes with the reflected light from the cavity with a difference in phase that depends on the deflection of the resonator (length of the cavity), and which changes with time as the resonator moves. The modulation of the reflected signal is maximum when the drumhead is actuated at its resonance frequency.

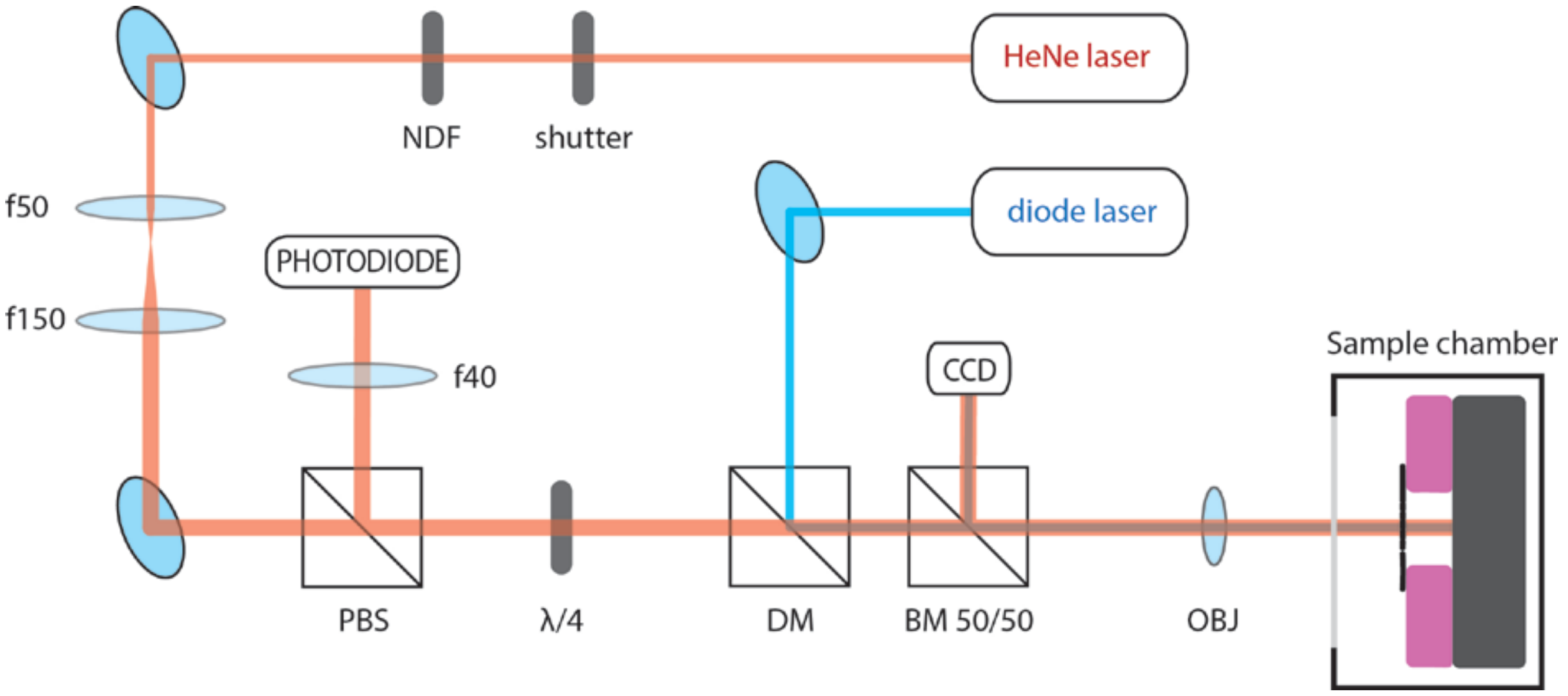

**Figure S-8**. Optical interferometer setup to measure the motion of $TaSe_2$ drumhead resonators. Abbreviations: NDF - neutral density filter, PBS - polarizing beam splitter, DM - dichroic mirror, BM - 50/50 beamsplitter, CCD - camera, OBJ - 50× objective lens.

Complete dataset

All the devices have a diameter of 3.2 µm, and the thickness uncertainty is 0.5 nm. The thickness for all the oxide devices is 3 nm smaller than the thickness of the pristine drum.

| Device | Thickness (nm) | Resonance Frequency $f_0$ (MHz) | | Quality Factor $Q$ | | $f$-$Q$ product ($10^{10}$ Hz) | |
|---|---|---|---|---|---|---|---|
| | | **Pristine** | **Oxide** | **Pristine** | **Oxide** | **Pristine** | **Oxide** |
| 39 | 6.2 | 7.7 | 47.1 | 60.7 | 73.8 | 0.05 | 0.35 |
| 42 | 7.3 | 15.5 | 34.8 | 221 | 1092 | 0.34 | 3.8 |
| 43 | 9.8 | 18.1 | 41.5 | 159 | 1035 | 0.29 | 4.29 |
| 36 | 12.6 | 13.9 | 31.1 | 91.7 | 1326 | 0.13 | 4.12 |
| 37* | 12.6 | 10.2 | 39.6 | 357 | 1058 | 0.36 | 4.19 |
| 32 | 14.9 | 7.1 | 41.7 | 105 | 747 | 0.07 | 3.12 |
| 29 | 15.8 | 20.1 | 41.8 | 315 | 732 | 0.63 | 3.06 |
| 30 | 15.8 | 19.4 | 39.7 | 260 | 736 | 0.51 | 2.92 |
| 41 | 18.9 | 18.4 | 43.8 | 162 | 1115 | 0.3 | 4.88 |
| 28 | 20.6 | 18.9 | 41.1 | 152 | 707 | 0.29 | 2.91 |
| 48 | 22.6 | 17.9 | 42.6 | 314 | 592 | 0.56 | 2.52 |
| 27 | 24.3 | 22.1 | 41.4 | 262 | 398 | 0.58 | 1.65 |
| 47 | 24.6 | 17.4 | 41.8 | 101 | 589 | 0.18 | 2.46 |
| 45 | 36.6 | 19.4 | 43.8 | 660 | 553 | 1.28 | 2.42 |
| 26 | 38.5 | 35.4 | 37.9 | 175 | 242 | 0.62 | 0.92 |
| 31 | 50.7 | 24.4 | 41.1 | 95 | 755 | 0.23 | 3.1 |
| 13 | 58.8 | 42.6 | 49.3 | 144 | 158 | 0.61 | 0.78 |
| 21 | 76.2 | 50.2 | 56.5 | 292 | 154 | 1.47 | 0.87 |
| 22 | 76.2 | 50.5 | 54.1 | 349 | 176 | 1.76 | 0.95 |
| 23 | 76.2 | 50.2 | 58.1 | 125 | 120 | 0.63 | 0.7 |
| 20 | 82.1 | 54 | 56.7 | 344 | 156 | 1.85 | 0.89 |
| 2 | 84.4 | 58.7 | 66.2 | 174 | 143 | 1.02 | 0.95 |
| 4 | 84.4 | 60.7 | 65.0 | 372 | 163 | 2.26 | 1.06 |
| 5 | 84.4 | 60.1 | 64.4 | 283 | 164 | 1.7 | 1.05 |
| 6 | 84.4 | 59.4 | 65.4 | 280 | 158 | 1.66 | 1.03 |
| 7 | 84.4 | 50.5 | 55.1 | 145 | 177 | 0.73 | 0.98 |
| 9 | 84.4 | 60.4 | 63.4 | 281 | 181 | 1.7 | 1.14 |
| 10 | 84.4 | 59.6 | 64.5 | 210 | 147 | 1.25 | 0.95 |
| 11 | 84.4 | 60.8 | 65.0 | 292 | 148 | 1.78 | 0.96 |
| 12 | 84.4 | 59.5 | 63.0 | 177 | 130 | 1.05 | 0.82 |
| 14 | 84.4 | 59.2 | 62.4 | 250 | 167 | 1.48 | 1.04 |
| 15 | 84.4 | 57.2 | 62.0 | 163 | 133 | 0.93 | 0.83 |
| 16 | 84.4 | 59.4 | 64.7 | 302 | 145 | 1.8 | 0.94 |
| 17 | 84.4 | 58.8 | 63.4 | 194 | 136 | 1.14 | 0.86 |
| 18 | 84.4 | 59.9 | 61.6 | 323 | 150 | 1.93 | 0.92 |
| 19 | 84.4 | 58.2 | 62.9 | 277 | 164 | 1.61 | 1.03 |
| 24 | 84.4 | 60.5 | 63.6 | 256 | 154 | 1.55 | 0.98 |
| 25 | 84.4 | 55.7 | 60.0 | 171 | 137 | 0.95 | 0.82 |
| 1 | 88.8 | 59.3 | 65.6 | 152 | 153 | 0.9 | 1.01 |
| 8 | 88.8 | 58.4 | 64.6 | 83 | 157 | 0.48 | 1.02 |
| 3 | 93.2 | 63 | 68.3 | 169 | 153 | 1.06 | 1.05 |

* Fig. 2

Effect of stress on the Q factor

In this section the effect of stress on the Q factor of the resonators is analyzed along the lines of reference [S9] in order to derive the thickness dependence of the ratio of the Q factors of laser oxidized and pristine $TaSe_2$ drums $Q_{ox}/Q_{pris}$. For a stress-free circular plate with mass $m$, spring constant $k_{plate}$ and Young's modulus $E = E_1 + iE_2$, it can be shown that the Q factor is given by $Q_{plate} = \frac{E_1}{E_2} = \frac{k_{plate,1}}{k_{plate,2}}$. In the presence of tensile stress, the real part of the spring constant will increase to a value $k_1 = k_{mem,1} + k_{plate,1}$ and the resonance frequency $f_0$ will become $f^2=k_1/m=f^2_{\mathrm{plate}}+f^2_{\mathrm{mem}}$. As discussed in the manuscript the Q factor will become:

$$Q=\frac{k_{\mathrm{mem,1}}+k_{\mathrm{plate,1}}}{k_{\mathrm{plate,2}}}=Q_{\mathrm{plate}}\frac{f^2}{f^2_{\mathrm{plate}}} \qquad \text{(S1)}$$

Based on the equations above and using $k_{\mathrm{plate,1,ox}}/k_{\mathrm{plate,1,pris}}=E_{1,\mathrm{ox}}/E_{1,\mathrm{pris}}$, it is now straightforward to derive the ratio of the Q factors of the oxidized and pristine drums:

$$\frac{Q_{ox}}{Q_{pris}}=\frac{f^2_{ox}}{f^2_{pris}}\frac{f^2_{plate,pris}}{f^2_{plate,ox}}\frac{E_{1,ox}}{E_{1,pris}}\frac{E_{2,pris}}{E_{2,ox}}=\frac{f^2_{ox}}{f^2_{pris}}\frac{m_{ox}}{m_{pris}}\frac{E_{2,pris}}{E_{2,ox}} \qquad \text{(S2)}$$

As evident from the data presented in the main text in the paper, this equation gives a good description of the thickness-dependence of the Q factor-ratio between the oxidized and pristine drums.

Wide frequency spectra of pristine and oxidized resonators

The spectra near the fundamental mechanical resonance frequency are used to study the effect of laser-oxidation on the mechanical properties of the drumheads. Figures S-9 and S-10 show the spectra of the drums over a wider frequency range. These spectra show that the lowest (fundamental) resonance frequency can be clearly identified as the peak with the highest intensity, in accordance with theory.

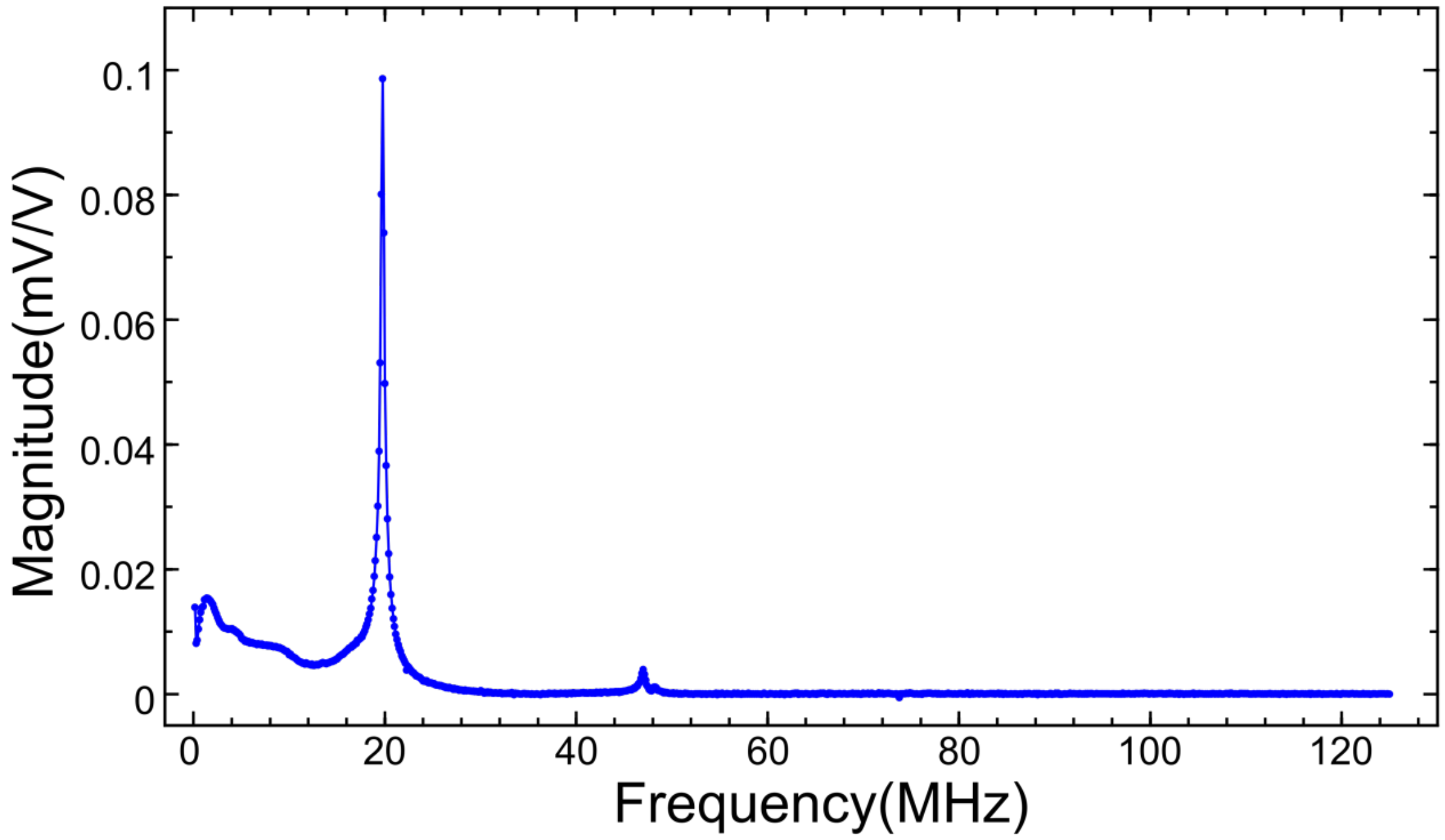

**Figure S-9.** Frequency spectrum of a $TaSe_2$ drumhead resonator.

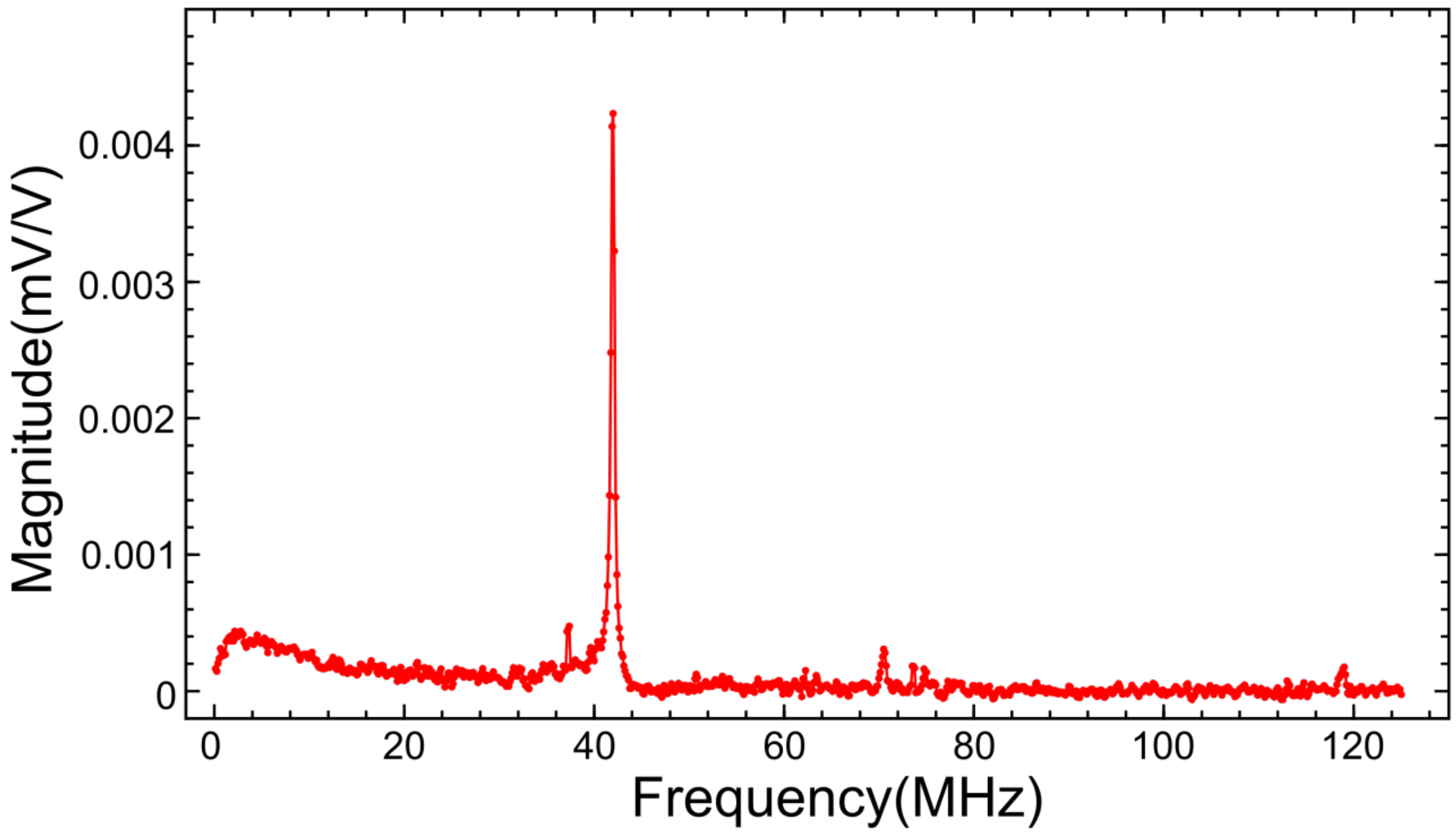

**Figure S-10.** Frequency spectrum of a laser-oxidized $TaSe_2$ drumhead resonator.

Local oxidation of resonators

The laser-oxidation technique is unique because it allows local oxidation of selected devices in contrast to other oxidation techniques that modify all exposed material. In Fig. S-11 we show the potential of the laser-oxidation technique for fabricating write-once-read-many (WORM) memory devices, where information is encoded in the resonance frequency of the Ta-based resonators.

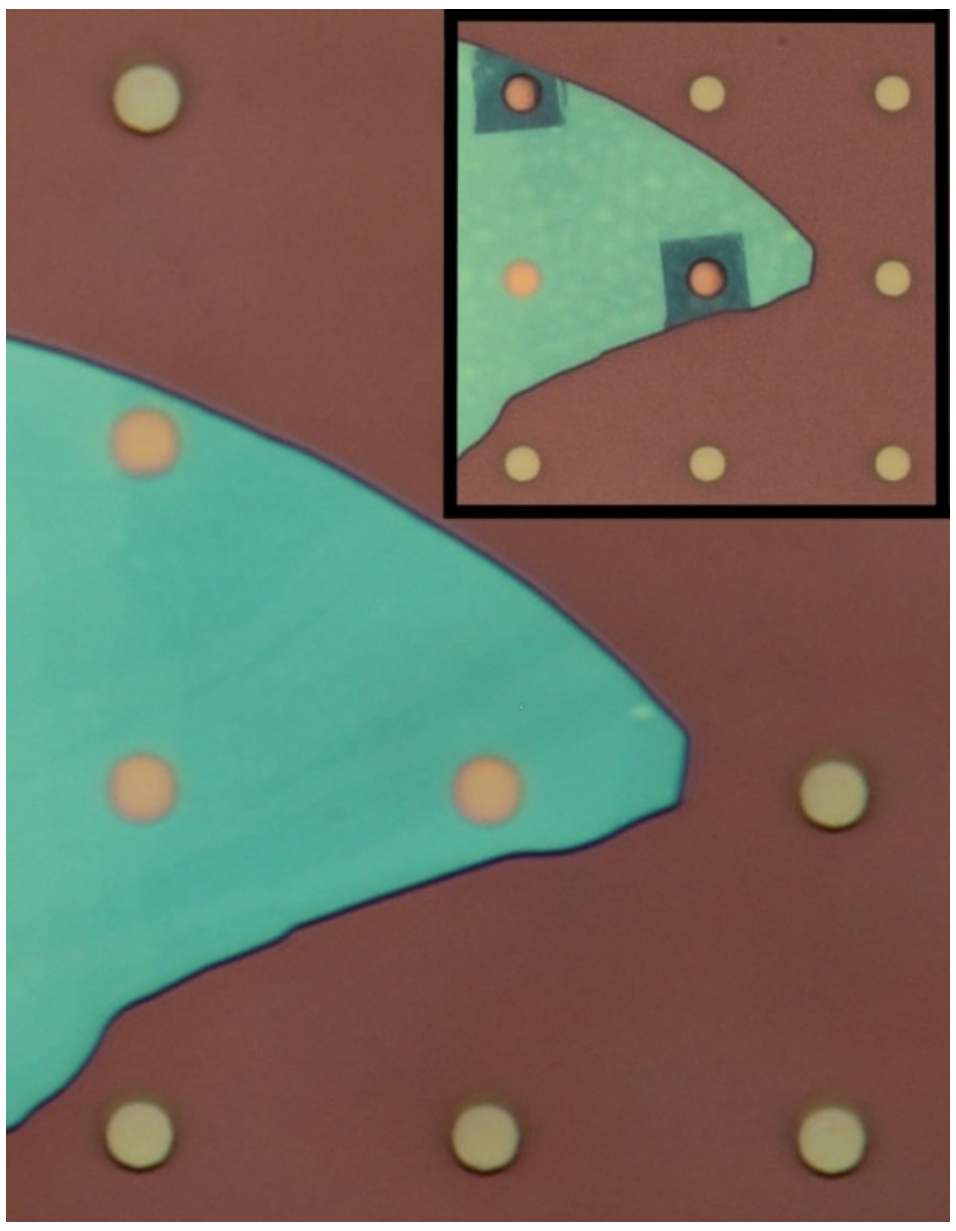

**Figure S-11.** Blank Ta-based mechanical memory. Inset: The laser oxidation technique is used to program the memory by writing bits. An interferometer can read the memory, giving a 1 or 0 depending on the measured resonance frequency of the drums.